\documentclass[10pt,conference,review]{IEEEtran}
\AtBeginDocument{%
  \providecommand\BibTeX{{%
    Bib\TeX}}}

\def\BibTeX{{\rm B\kern-.05em{\sc i\kern-.025em b}\kern-.08em
    T\kern-.1667em\lower.7ex\hbox{E}\kern-.125emX}}
\usepackage[utf8]{inputenc}
\usepackage{graphicx}     % Add graphics capabilities
\usepackage{amsmath}      % Better maths support
\usepackage{eurosym}
\usepackage{paralist}
\usepackage{mdframed}
\usepackage{blindtext}
\usepackage{soul}
\usepackage{comment}
\usepackage[TS1,T1]{fontenc}

\usepackage{enumitem}
\usepackage{enumitem}
\usepackage{multirow}
\usepackage[labelfont=bf]{caption}

\newenvironment{itemize*}%
  {\begin{itemize}%
    \setlength{\itemsep}{0pt}%
    \setlength{\parskip}{0pt}}%
  {\end{itemize}}
  
\usepackage{array}
\usepackage{longtable}

\newcolumntype{L}[1]{>{\raggedright\let\newline\\\arraybackslash\hspace{0pt}}m{#1}}
\newcolumntype{C}[1]{>{\centering\let\newline\\\arraybackslash\hspace{0pt}}m{#1}}
\newcolumntype{R}[1]{>{\raggedleft\let\newline\\\arraybackslash\hspace{0pt}}m{#1}}

\usepackage{enumitem}

\usepackage{hyperref}
\AtBeginDocument{%

}

\usepackage[table]{xcolor}
\usepackage[dvipsnames]{xcolor}
\usepackage{balance}
\usepackage[nomain,acronym]{glossaries}
\makeglossaries

\newcommand{\abkuerzung}[2]{\newacronym{#1}{#1}{#2}}

\abkuerzung{AADL}{Architecture Analysis \& Design Language}

\abkuerzung{ADL}{architecture description language}

\abkuerzung{ADR}{Action Design Research}

\abkuerzung{API}{application programming interface}

\abkuerzung{AS}{abstract syntax}

\abkuerzung{AST}{abstract syntax tree}

\abkuerzung{CBSE}{component-based software engineering}

\abkuerzung{CD}{Class Diagram}

\abkuerzung{CFG}{context-free grammar}

\abkuerzung{CoCo}{context condition}

\abkuerzung{CS}{concrete syntax}

\abkuerzung{CPS}{cyber-physical system}

\abkuerzung{DoE}{design of experiments}

\abkuerzung{DSL}{domain-specific language}

\abkuerzung{FSM}{finite-state machine}

\abkuerzung{GPL}{general-purpose programming language}

\abkuerzung{KR}{knowledge representation}

\abkuerzung{LCDP}{low-code development platform}

\abkuerzung{DPL}{DSL product line}

\abkuerzung{M2M}{model-to-model}

\abkuerzung{M2T}{model-to-text}

\abkuerzung{MBSE}{model-based systems engineering}

\abkuerzung{MDE}{model-driven engineering}

\abkuerzung{MDSE}{model-driven systems engineering}

\abkuerzung{OCL}{Object Constraint Language}

\abkuerzung{SC}{statechart}

\abkuerzung{Sem}{semantics}

\abkuerzung{SME}{small and medium-sized enterprise}

\abkuerzung{SLE}{software language engineering}

\abkuerzung{SOS}{Structural Operational Semantics}

\abkuerzung{SPL}{software product line}

\abkuerzung{SysML}{Systems Modeling Language}

\abkuerzung{TS}{technological space}

\abkuerzung{TSCA}{time-synchronous channel automata}

\abkuerzung{TSPA}{time-synchronous port automata}

\abkuerzung{UML}{Unified Modeling Language}

\usepackage{xspace}
\usepackage{ifdraft}

\newcommand*{\ie}{\textit{i.e.,}\@\xspace}
\newcommand*{\eg}{\textit{e.g.,}\@\xspace}
\newcommand*{\cf}{\textit{cf.}\@\xspace}

\makeatletter
\newcommand*{\etc}{%
  \@ifnextchar{.}%
  {\textit{etc}}%
  {\textit{etc.}\@\xspace}%
}
\makeatother

\definecolor{se-green}{RGB}{0,128,0}
\definecolor{se-blue} {RGB}{0,0,204}

\newcounter{papernumber}

\newcounter{requirement}[section]

\definecolor{MyBoxText}{RGB}{255,255,255}
\definecolor{MyBoxBG}{RGB}{13,50,153}

\newboolean{edit}
\setboolean{edit}{false}

\ifthenelse{\boolean{edit}}{%
  \renewcommand{\baselinestretch}{2} 
}{%
}

\usepackage{ifdraft}
\usepackage{ifthen}

\newboolean{debug} 
\setboolean{debug}{false}
 
\ifthenelse{\boolean{debug}}{
  \usepackage{showframe} %% activate frames 
  \usepackage{todonotes} %% activate todos
  \newcommand{\xynote}[2]{\todo[inline]{#1: #2}}
  
  \newcommand{\smallnote}[2]{
    \par\noindent\makebox[\textwidth][c]{%
      \fbox{
        \begin{minipage}{0.9\textwidth}
          \scriptsize{\color{#1}{#2}}
        \end{minipage}
      }
    }
  }
  \newcommand{\hint}[1]{\smallnote{black}{#1}}
  \newcommand{\thought}[1]{\smallnote{teal}{#1}}
  \newcommand{\actionitem}[2]{\noindent$\circ$ #1 (#2)\\}
  \newcommand{\xactionitem}[2]{\noindent$\times$ #1 (#2)\\}
  
  \newcommand{\del}[1]{\textcolor{red}{\sout{#1}}} % please delete
}{%
  \usepackage[disable]{todonotes}
  \hypersetup{hidelinks} 
  \newcommand{\xynote}[2]{}
  \newcommand{\smallnote}[1]{}
  
  \newcommand{\hint}[1]{}
  \newcommand{\thought}[1]{}
  \newcommand{\actionitem}[2]{}
  \newcommand{\xactionitem}[2]{}
  
  \newcommand{\del}[1]{} % please delete
}

\DeclareUnicodeCharacter{301}{XXXXXXXXXXX}
\DeclareUnicodeCharacter{302}{XXXXXXXXXXX}
\DeclareUnicodeCharacter{308}{XXXXXXXXXXX}

\usepackage{listings}
\newenvironment{ownfigure}[0]%
{\begin{figure}[htb!]}%\stepcounter{table}}%
{\end{figure}}

\usepackage{color}
\definecolor{DarkRed}{rgb}{0.75,0,0}
\definecolor{Lightgreen}{rgb}{0.588,1.0,0.588}
\definecolor{DarkGreen}{rgb}{0,0.5,0}
    
\lstdefinelanguage{MontiArc}[]{Java}{
  morekeywords={component, port, in, out, inv, package, import, connect, autoconnect}
}

\lstdefinelanguage{myJava}[]{Java}{
  commentstyle=\color{DarkGreen}\itshape 
}

\lstdefinelanguage{MontiArcAutomaton}[]{Java}{
  morekeywords={component, port, in, out, inv, package, import, connect,
  autoconnect, automaton, state, ocl, java, initial, final,
  noCompletion, chaosCompletion, var, mode, activate, transitions,
  modetransitions}, commentstyle=\color{DarkGreen}\itshape }

\lstdefinelanguage{MCConfig} { 
    morekeywords={config, Require, Model} 
}

\lstdefinelanguage{Manifest} { 
    morekeywords={Manifest, Bundle, ManifestVersion, Name, SymbolicName,
      Version, Require
    } 
}

\lstdefinelanguage{mcGrammar}[]{}{
  morekeywords={
    grammar, package, path, parser, lexer, nows, noslcomments, nomlcomments, 
    noident, nostring, noanything, nocharvocabulary, dotident, identrule,
    xmlcomments, hashcomments, texcomments, freemarkercomments, concept, 
    globalnaming, define, usage, options, true, false, protected, ident, 
    compilationunit
  }
}

\lstdefinelanguage{mcLng}[]{}{
  morekeywords={
    dsltool, language, package, path, parser, root, parsingworkflow, 
    rootfactory, lexer, nows, noslcomments, nomlcomments, noident, nostring,
    dotident, concept, globalnaming, define, usage, options, true, false, 
    protected, ident
  }
}

\lstdefinelanguage{mcManifest}[]{}{
  morekeywords={
    bundle, Bundle, Name, SymbolicName, true, false, Main, Class, 
    Version, Activator, Localization, Require, 
    Exclude, Eclipse, LazyStart, Vendor, Export, Package, 
    ClassPath
  }
}

\lstdefinelanguage{Alloy}[]{Java}{
commentstyle=\color{DarkGreen}\itshape,
  morekeywords={abstract,sig,->,fact,pred,fun,run,for,iff,
  not,no,one,all,some,lone,\#,set,in,and,or,but,exactly,none,univ,Int,assert,check},
  otherkeywords = {[2]????},
    morekeywords = {[2]????},
    keywordstyle={[2]\color{blue}},
    otherkeywords = {[3]????,<,<->,->, &, |, =, !=, !,<:,~},
    morekeywords = {[3]????,<,<->,->, &, |, =, !=, !,<:,~},
    keywordstyle={[3]\color{blue}}
  }

\lstdefinelanguage{mccd}[]{Java}{
  morekeywords={classdiagram,abstract,<<singleton>>,class,int,String,
  association,composition,extends}
}

\lstdefinelanguage{FreeMarker}[]{}{
  keywordsprefix={\#},
  keywords={in},
  commentstyle=\color{DarkGreen}\itshape }

\lstdefinelanguage{Mona}[]{}{
  morekeywords={ex0,all0,ex1,all1,ex2,all2,var0,var1,var2,pred,in,notin,include,union,inter,empty,assert},
  morecomment=[l]{\#},
  commentstyle=\color{DarkGreen}\itshape,
  otherkeywords = {[2]????,next,boolean,init,case,esac},
  morekeywords = {[2]????,next,boolean,init,case,esac},
  otherkeywords = {[3]????,<,<=>,=>, &, |, =, !=, !},
  morekeywords = {[3]????,<,<=>,=>, &, |, =, !=, !},
}

\lstdefinelanguage{myPython}[]{Python}{
  morekeywords={assert},
  morecomment=[l]{\#},
  commentstyle=\color{DarkGreen}\itshape,
}

\lstdefinelanguage{GeneratorConfiguration}[]{Java} {
  morekeywords={
    template, 
    generator, 
    ast, 
    runtime},
}

\lstdefinelanguage{ApplicationConfiguration}[]{Java} {
  morekeywords={
    application,
    behaviors,
    bindings,
    classdiagrams,
    components,
    factories,
    generators,
    map,
    to},
}

\lstdefinelanguage{Isabelle}[]{} {
    morekeywords={
        datatype,
        typedef},
}

\definecolor{TableHead}{HTML}{404040}

\newcommand{\rowHeader}[1]{\cellcolor{#1} #1}

\definecolor{Plan}{HTML}{5BEFC5}
\newcommand{\pplan}[0]{\rowHeader{Plan}}

\definecolor{Code}{HTML}{5AADFF}
\newcommand{\pcode}[0]{\rowHeader{Code}}

\definecolor{Build}{HTML}{BDBBF5}
\newcommand{\pbuild}[0]{\rowHeader{Build}}

\definecolor{Test}{HTML}{c381e2}
\newcommand{\ptest}[0]{\rowHeader{Test}}

\definecolor{Release}{HTML}{ff858e}

\definecolor{Deploy}{HTML}{FBCCC0}
\newcommand{\pdeploy}[0]{\rowHeader{Deploy}}

\definecolor{Operate}{HTML}{ffd685}
\newcommand{\poperate}[0]{\rowHeader{Operate}}

\definecolor{Monitor}{HTML}{C8FF88}
\newcommand{\pmonitor}[0]{\rowHeader{Monitor}}

\renewcommand{\tablename}{Tbl.}
\renewcommand{\thetable}{\arabic{table}}

\begin{document}
% kel: kp ob das hier die richtige stelle für den command ist haha
\raggedbottom
\title{Digital Twins for Software Engineering Processes}

\author{
    \IEEEauthorblockN{Robin Kimmel$^{1}$,  Judith Michael$^{2}$, 
    Andreas Wortmann$^{1}$,
    Jingxi Zhang$^{1}$}
    
    \IEEEauthorblockA{  
    $^{1}$
    \textit{ISW, University of Stuttgart, Stuttgart, Germany}, {firstname.lastname}@isw.uni-stuttgart.de
    }
    \IEEEauthorblockA{  
    $^{2}$
    \textit{SE, RWTH Aachen University, Aachen, Germany}, 
    michael@se-rwth.de
    }
}
\maketitle
\begin{abstract}
Digital twins promise a better understanding and use of complex systems.
To this end, they represent these systems at their runtime and may interact with them to control their processes.
Software engineering is a wicked challenge in which stakeholders from many domains collaborate to produce software artifacts together.
In the presence of skilled software engineer shortage, our vision is to leverage DTs as means for better representing, understanding, and optimizing software engineering processes to (i) enable software experts making the best use of their time and (ii) support domain experts in producing high-quality software. 
This paper outlines why this would be beneficial, what such a digital twin could look like, and what is missing for realizing and deploying software engineering digital twins. 
\end{abstract}
%\begin{CCSXML}
%<ccs2012>
%  <concept_id>10010520.10010553.10010562</concept_id>
%  <concept_desc>Computer systems organization~Embedded systems</concept_desc>
%  <concept_significance>500</concept_significance>
%  <concept_id>10010520.10010575.10010755</concept_id>
%  <concept_desc>Computer systems organization~Redundancy</concept_desc>
%  <concept_significance>300</concept_significance>
%  <concept_id>10010520.10010553.10010554</concept_id>
%  <concept_desc>Computer systems organization~Robotics</concept_desc>
%  <concept_significance>100</concept_significance>
%  <concept_id>10003033.10003083.10003095</concept_id>
%  <concept_desc>Networks~Network reliability</concept_desc>
%  <concept_significance>100</concept_significance>
%</ccs2012>
%\end{CCSXML}

%\ccsdesc[500]{Computer systems organization~Embedded systems}


\section{Motivation} 

The shortage~\cite{FinancialTimesWorkerShortage} of skilled workers in software engineering is a growing concern with significant implications for all industries depending on software. 
As the amount of software around us and the demand for novel software solutions continues to rise, the growth of the workforce of qualified software engineers is not keeping pace.
With fewer skilled engineers available, companies struggle to develop new and advanced software solutions, and the pace of innovation slows~\cite{ForbesWorkerShortage}, potentially leading to stagnation in technological advancements.
This gap between required and provided software engineering capacities prevents growth and stifles innovation.

Computer science has always been a story of increasing abstraction and improving automation. 
Automation of software engineering processes can help bridge the gap caused by the shortage of skilled workers by taking over well-understood, mundane, repetitive tasks, allowing software engineers to focus on more complex and creative challenges.
Moreover, the rise of AI-support for all phases of the software development and operations (DevOps) life-cycle suggests that an increasingly comprehensive digital representation of relevant aspects of the DevOps activities becomes feasible and can be exploited to (support) automated decision-making about the process. 
Such a representation could be considered a digital twin (DT) of a software engineering process.

Based on the understanding of DTs based on the data flow model of DTs~\cite{kritzinger2018digital} and the 5D model of DT components~\cite{tao2018digital}, we envision that such DTs, which can sense data from the represented system (\ie a software engineering process), reason about this data using internal models, and act on the represented system (\eg to suggest code improvements) will become holistic software engineering co-pilots that encompass the complete DevOps cycle instead of living in the IDE only.

Next, \autoref{sec:Background} illustrates what we understand as DTs, before \autoref{sec:DTsofSPs} outlines the idea of DTs of software engineering processes.
Afterward, \autoref{sec:Example} illustrates an example application before \autoref{sec:Discussion} discusses our idea.
Then, \autoref{sec:FuturePlans} lays out our next steps toward this idea and \autoref{sec:Conclusion} concludes.

\section{Background}
\label{sec:Background}

\paragraph*{Digital Twins} The most prominent definitions of DTs (a) separate them from digital models and digital shadows by requiring automated data flows from the original system (to sense changes in it) and to the original system (to induce changes to it)~\cite{kritzinger2018digital} and (b) require that they comprise five dimensions~\cite{tao2018digital}: 
\begin{inparaenum}[(1)]
\item an interface to the original system,
\item data from and about that system,
\item models representing (parts of) that system,
\item services producing added value based on data and models, and 
\item connections between all of these.
\end{inparaenum}

\begin{figure}[h]
    \centering
    \includegraphics[width=\linewidth]{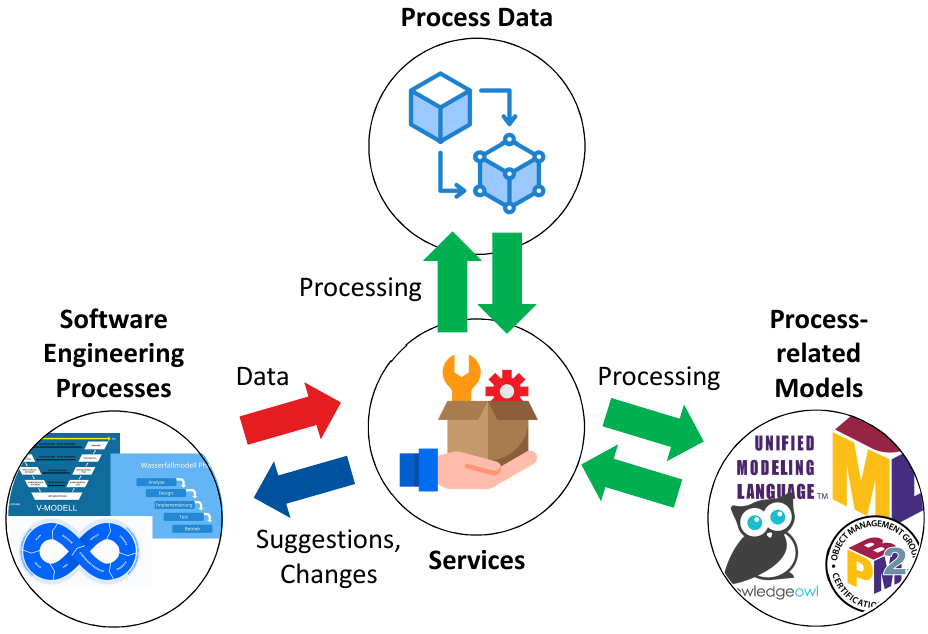}
    \caption{5D DTs applied to software processes~\cite{tao2018digital}}
    \label{fig:devops_dt_overview}
\end{figure}

Consequently, a DT is a software system that connects to the original system, observes data from it, uses its models about that system and the services to reason about these observations, and suggests changes to the system--
which can be software themselves~\cite{lai2023wirelessdt,heluany2024interplay,ahlgren2021facebook}
And while there is a tremendous body of work on DTs, such as mapping studies on their nature~\cite{dalibor2022cross}, analyses of their characteristics~\cite{jones2020characterising}, surveys on their reference architectures~\cite{ferko2023standardisation}, systematic literature reviews on the development of DT software in application areas~\cite{guinea2024digital} and plethora of publications on their application to specific challenges, a DTs of software engineering processes is missing.

\paragraph*{Automated Software Engineering Pipelines} 
There are many for automating software engineering, such as SonarQube \cite{SonarQube}, GitLab's continuous integration and continuous delivery (CI/CD) \cite{GitLab}, or Atlassian Bamboo \cite{Bamboo}, which primarily are used to analyze the code base and only after committing changes to the code base to the server. 
Some are integrated into comprehensive tool suites, where planning can be conducted through issue boards and the testing and deployment of software can be automated through the CI/CD pipeline. 
Although these tools facilitate the tracking of development processes, optimization techniques, such as work splitting and the simulation of software development processes, the observation of the running software products and their links to the development artifacts (not limited to the code base) are still largely underexplored.
While there is ongoing research that investigates leveraging DTs on the operations side of DevOps~\cite{combemale2023model}, a holistic DT of software engineering processes that encompasses all relevant artifacts and engineering phases does not exist yet.
\section{Dimensions of Software Process Twins}
\label{sec:DTsofSPs}

With software process DTs, we suggest transferring a part of tool and automation control from stakeholders to a comprehensive DT that (a) encompasses, and relates, all relevant activities of a specific software engineering process, and (b) actively suggests or realizes changes to the process and its artifacts instead of reacting on stakeholder inputs only.

%%%%%%%%%%%%%%%%%%%%%%%%%%%%%%%%%%%%%%%%%%%%%%%%%%%%%%%%%%%%%%%%%%%%%%%%%%%%%%%%
% 5D Model in Theory
%%%%%%%%%%%%%%%%%%%%%%%%%%%%%%%%%%%%%%%%%%%%%%%%%%%%%%%%%%%%%%%%%%%%%%%%%%%%%%%%

According to the 5D model of DTs (\cf \autoref{fig:devops_dt_overview}), such twins need
\begin{inparaenum}[(1)]
\item an interface to obtain data from the development process,
\item data from that process,
\item models about the process (the \emph{knowledge}),
\item services computing changes and suggestions (the \emph{reasoning}) on the process and its artifacts, and
\item connections between all of them.
\end{inparaenum}
Therefore, we assume that the services are units of computation such as twins and that they connect data obtained via the interface to the process to internal models about it before producing actions regarding the process.
The following outlines these dimensions for a DT for software engineering processes.

%%%%%%%%%%%%%%%%%%%%%%%%%%%%%%%%%%%%%%%%%%%%%%%%%%%%%%%%%%%%%%%%%%%%%%%%%%%%%%%%
% Data dimension
%%%%%%%%%%%%%%%%%%%%%%%%%%%%%%%%%%%%%%%%%%%%%%%%%%%%%%%%%%%%%%%%%%%%%%%%%%%%%%%%

Interfaces to the process come in many forms and can include anything digitally observable in current and future SE processes. 
As of today, this includes data from   
\begin{inparaenum}[(1)]
    \item project management tools (such as issue tickets, merge requests, \etc),
    \item interactions with CI/CD tools, co-Pilots, test results, 
    \item observing IDEs and the artifacts (code, dependency management, debugging information) being manipulated therein. 
\end{inparaenum}
Aside from these obvious data sources, further data might support a better understanding while improving the SE process as well, such as 
\begin{inparaenum}[(1)]
    \setcounter{enumi}{3}
    \item information about the availability of developers, \eg from their work calendars (to automatically plan the distribution of tasks and monitor workload conflicts), or 
    \item conversations, such as meeting protocols or chat logs, about the current project and related projects (to identify topics that have been problematic in the past and might raise issues again),
\end{inparaenum}

%%%%%%%%%%%%%%%%%%%%%%%%%%%%%%%%%%%%%%%%%%%%%%%%%%%%%%%%%%%%%%%%%%%%%%%%%%%%%%%%
% Model dimension
%%%%%%%%%%%%%%%%%%%%%%%%%%%%%%%%%%%%%%%%%%%%%%%%%%%%%%%%%%%%%%%%%%%%%%%%%%%%%%%%

Models of a software engineering process include process models that describe graphs of activities and relate these to involved stakeholders, data models about development information (e.g., tickets, properties of developers), constraint models (e.g., about available resources) that the services of the DTs use, as well as AI models, to make decisions about the process. 
In addition, models created within a software engineering process should be considered, e.g., architecture design models, goal models for specifying system requirements from the user perspective, sequence models specifying the system behavior, or test models describing test cases and test data.

%%%%%%%%%%%%%%%%%%%%%%%%%%%%%%%%%%%%%%%%%%%%%%%%%%%%%%%%%%%%%%%%%%%%%%%%%%%%%%%%
% Service dimension
%%%%%%%%%%%%%%%%%%%%%%%%%%%%%%%%%%%%%%%%%%%%%%%%%%%%%%%%%%%%%%%%%%%%%%%%%%%%%%%%

\begin{figure}[h]
    \centering
    \includegraphics[width=0.8\linewidth]{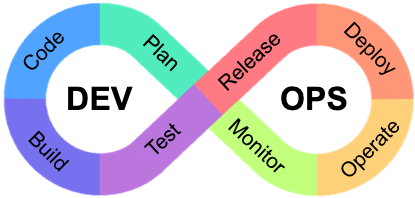}
    \caption{The eight phases of DevOps}
    \label{fig:DevOps}
\end{figure}

Based on the data and models, services can represent the state of the process, as well as analyze it and propose change suggestions.
Various technologies for such services already exist and can become part of the software process DTs easily.
\autoref{tbl:EVA} lists examples of such services for the different software engineering activities based on the DevOps process model (cf. \autoref{fig:DevOps}). 
These activities are partially aligned with the DevOps cycle to encompass both the development and the operations phases. However, as the underlying process model, any other model, for example, the V-Model and the Waterfall model, can be used.
For each service, we summarize inputs (data and models), its reasoning activities, and potential outputs. 

% \kel{Bin nicht so ganz zufrieden damit wie die Fraben aussehen}
% \aw{Ich ebenfalls nicht. Zum Glück kann man in Latex eigene Farben definieren. Versuche es mal mit den Farben aus \url{https://dinext-group.com/wp-content/uploads/2023/03/DevOps-1.png} und schau im Latex mal nach der Definition von "pplan"}

% \aw{Bitte übrigens jede Zeile einer einzigen Phase zuordnen. Auch wenn es in mehreren Anwendbar wäre (im Text sowas schreiben wie: "jeweils der passendsten Phase zugeordnet")}

% \aw{Bitte zu möglichst vielen Services Quellen angeben (in Spalte Processing)}

% \aw{Idealerweise gibt es zu jeder Phase mindestens zwei Zeile in der Tabelle (wir haben ja noch platz)}

% EVA Tabelle: siehe https://docs.google.com/spreadsheets/d/ 1XcuTBSCkKYA1eSd2Hx0fL048Ia9G9QLknYNO5jmWnJg/edit?usp=sharing las 
% für weitere Kommentare und Infos zu den Einträgen
\bgroup
\def\arraystretch{1.3} % padding
\begin{table*}[t] % double column and floating top
\caption{Service input, processing, and output for selected software engineering twin activities, based on different DevOps phases \cite{ebert2016devops}. The comprehensive list can be found here \cite{Drive}
}
\centering
\begin{tabular}{|p{1.2cm} p{4cm} p{5.5cm} p{5.5cm}|}
\hline
%%%%%%%%%%%%%%%%%%%%%%%%%%%%%%%%%%%%%%%%
% Header
\textbf{Phase} & 
\textbf{Service Input (Data \& Models)} & 
\textbf{Service Reasoning} & 
\textbf{Service Output (Suggestions, Changes)} \\
\hline
%%%%%%%%%%%%%%%%%%%%%%%%%%%%%%%%%%%%%%%%
% Row 

\pplan & 
\cellcolor{Plan} Open issues and successfully closed issues. & 
\cellcolor{Plan} Calculation of similarity between issues closed by Developer X and currently open issues \cite{tunio2018crowdsourcing} \cite{dam2019towards}. & 
\cellcolor{Plan} Proposed ticket assignments based on similar, successfully resolved tickets from the past. \\  
\hline
% %%%%%%%%%%%%%%%%%%%%%%%%%%%%%%%%%%%%%%%%
% % Row 
% \pplan & 
% \cellcolor{Plan} Developer X's assigned issues with estimated effort of N hours for the next S-week sprint. Continuous monitoring of new appointments in X's calendar and tickets completed by X. & 
% \cellcolor{Plan} Calculation of whether sufficient time remains until the end of S weeks to address remaining open tickets. & 
% \cellcolor{Plan} Warning to developer and/or team leader if available time is significantly underestimated. \\
% \hline
%%%%%%%%%%%%%%%%%%%%%%%%%%%%%%%%%%%%%%%%
% Row 
\pplan & 
\cellcolor{Plan} Newly created requirements (Frontend or Backend). & 
\cellcolor{Plan} Adaptation of requirements to project standard with developer specific views. & 
\cellcolor{Plan} Output of requirements in various formats such as natural language, UML, etc. with user specific preferences. Include recommendations for code snippets, UI layouts etc. \\
\hline
% %%%%%%%%%%%%%%%%%%%%%%%%%%%%%%%%%%%%%%%%
% % Row 
% \pcode & 
% \cellcolor{Code} Architecture description (or models) and code currently under development. & 
% \cellcolor{Code} Analysis of code for potential contradictions with the architecture description (e.g., logically separate components unexpectedly communicating). & 
% \cellcolor{Code} Indication of contradictory elements in the architecture document and identification of relevant stakeholders. \\
%%%%%%%%%%%%%%%%%%%%%%%%%%%%%%%%%%%%%%%%
% Row 
\pcode & 
\cellcolor{Code} Declarative knowledge about the project, for example, facts about existing race conditions (two or more threads accessing a specific variable). & 
\cellcolor{Code} Store information and provide an interface for a developer to retrieve declarative knowledge while working inside an IDE \cite{leung2023automated}. & 
\cellcolor{Code} Provide the developer with declarative knowledge about the project, ticket etc. or directly recommend snippets based on the knowledge. \\
\hline
%%%%%%%%%%%%%%%%%%%%%%%%%%%%%%%%%%%%%%%%
% Row 
\pcode & 
\cellcolor{Code} Currently created code from an IDE. Developer profile. & 
\cellcolor{Code} Comparison via embeddings with a database of code snippets \cite{bader2021ai}. & 
\cellcolor{Code} Indication in the IDE whether similar code has been implemented elsewhere or how others have implemented it. \\
\hline
%%%%%%%%%%%%%%%%%%%%%%%%%%%%%%%%%%%%%%%%
% Row 
\pcode & 
\cellcolor{Code} A developer trying to create a new feature in an existing code base. & 
\cellcolor{Code} Using a model trained via deep learning from GitHub commits to emulate, attempts to improve readability \cite{vitale2023using}. & 
\cellcolor{Code} Provide an improved version of the relevant code with better readability to allow the developer to make better decisions more easily. \\
\hline
%%%%%%%%%%%%%%%%%%%%%%%%%%%%%%%%%%%%%%%%
% Row 
\pcode & 
\cellcolor{Code} Currently created code from an IDE. & 
\cellcolor{Code} Extract and analyze dependencies between different artifacts \cite{greifenberg2017towards}. & 
\cellcolor{Code} Suggestions on how to structure incremental builds or for reducing coupling between components. \\
\hline
%%%%%%%%%%%%%%%%%%%%%%%%%%%%%%%%%%%%%%%%
% Row 
\pbuild & 
\cellcolor{Build} Changes in dependency versions in build artifacts (e.g., Maven's POM). & 
\cellcolor{Build} Analysis of issues (both on GitHub and internal code base) that arose after the version upgrade. Identification of code changes that match patterns corresponding to known design patterns in the local code base. & 
\cellcolor{Build} Suggestions for automatic local code adaptations based on identified patterns in related issues. Indications of potentially affected areas based on issue patterns. Recommendations for updating other dependencies based on changes observed in issue-related code. \\ 
\hline
%%%%%%%%%%%%%%%%%%%%%%%%%%%%%%%%%%%%%%%%
% Row 
\pbuild & 
\cellcolor{Build} Code and target hardware specifications. & 
\cellcolor{Build} Estimation of energy consumption on target hardware. & 
\cellcolor{Build} Estimated consumption of code sections, categorized as green, yellow, or red. Recommendations for improvements. \\
\hline
% %%%%%%%%%%%%%%%%%%%%%%%%%%%%%%%%%%%%%%%%
% % Row 
% Code & Code and target hardware specifications. & Estimation of energy consumption on target hardware. & Estimated consumption of code sections, categorized as green, yellow, or red. Recommendations for improvements. \\
% \hline
% %%%%%%%%%%%%%%%%%%%%%%%%%%%%%%%%%%%%%%%%
% % Row 
% Plan \& Code \& Build \& Test & All inputs and queries related to the development process in natural language, including documentation, meetings, and records of design decisions. & Construction of a business process model, evaluation of queries against the process model, and translation of results back into natural language. & Responses to queries about the (ongoing) process in natural language. \\
% \hline
%%%%%%%%%%%%%%%%%%%%%%%%%%%%%%%%%%%%%%%%
% Row 
\ptest & 
\cellcolor{Test} Code changes in the IDE. &
\cellcolor{Test} Automatic creation and execution of tests in real-time to anticipate potential errors \cite{reddy2020quickly} \cite{fatima2022flakify}. & 
\cellcolor{Test} Immediate feedback for the developer.\\
\hline
%%%%%%%%%%%%%%%%%%%%%%%%%%%%%%%%%%%%%%%%
% Row 
\ptest & 
\cellcolor{Test} Existing code base of a micro-service architecture as containerized applications. & 
\cellcolor{Test}
Execution of a combination of different security scanners, such as for example docker-bench-security, nmap, terrascan  \cite{unver2023automatic}. & 
\cellcolor{Test} Security report for the given set of applications derived from different tools. \\
\hline
% %%%%%%%%%%%%%%%%%%%%%%%%%%%%%%%%%%%%%%%%
% % Row 
% Plan Code/Build & System requirements and code. & Application of automatic requirements tracing techniques. & Verification (OK/NOK) that requirements are fulfilled in the implementation. \\
% \hline
% %%%%%%%%%%%%%%%%%%%%%%%%%%%%%%%%%%%%%%%%
% % Row 
% Plan Code/Build & System requirements and code. & Identification of unnecessary components. & Report on unneeded components. \\
% \hline
% %%%%%%%%%%%%%%%%%%%%%%%%%%%%%%%%%%%%%%%%
% % Row 
% \prelease & 
% \cellcolor{Release} TODO & 
% \cellcolor{Release} TODO & 
% \cellcolor{Release} TODO \\
% \hline
% %%%%%%%%%%%%%%%%%%%%%%%%%%%%%%%%%%%%%%%%
% % Row 
% \prelease & 
% \cellcolor{Release} TODO & 
% \cellcolor{Release} TODO & 
% \cellcolor{Release} TODO \\
% \hline
%%%%%%%%%%%%%%%%%%%%%%%%%%%%%%%%%%%%%%%%
% Row 
\pdeploy & 
\cellcolor{Deploy} Code in development, models of target platform and deployments.& 
\cellcolor{Deploy} Analysis of compatibility of code to variants of target platforms and deployments \cite{kirchhof2022model}. & 
\cellcolor{Deploy} Validity of potential code deployment or list of incompatible software parts.\\
\hline
% %%%%%%%%%%%%%%%%%%%%%%%%%%%%%%%%%%%%%%%%
% % Row 
% \pdeploy & 
% \cellcolor{Deploy} TODO & 
% \cellcolor{Deploy} TODO & 
% \cellcolor{Deploy} TODO \\
% \hline
%%%%%%%%%%%%%%%%%%%%%%%%%%%%%%%%%%%%%%%%
% Row 
\poperate & 
\cellcolor{Operate} Logged system operation events. & 
\cellcolor{Operate} Analyze runtime behavior \cite{bhagwan2021learning}. & 
\cellcolor{Operate} Suggestions on how to alter the system configuration of dynamically adaptive systems. \\
\hline
%%%%%%%%%%%%%%%%%%%%%%%%%%%%%%%%%%%%%%%%
% Row 
\poperate & 
\cellcolor{Operate} Current real time energy consumption. & 
\cellcolor{Operate} Peak analysis or comparison to predefined energy goals. & 
\cellcolor{Operate} Show times of highest energy need to readjust system configuration or create energy usage reports. \\
\hline
%%%%%%%%%%%%%%%%%%%%%%%%%%%%%%%%%%%%%%%%
% Row 
\pmonitor & 
\cellcolor{Monitor} Sustainability goals as non-functional requirements and system monitoring data. & 
\cellcolor{Monitor} Calculation of metrics and comparison with sustainability targets. & 
\cellcolor{Monitor} Report on achieved or failed sustainability goals. \\
\hline
%%%%%%%%%%%%%%%%%%%%%%%%%%%%%%%%%%%%%%%%
% % Row 
% \pmonitor & 
% \cellcolor{Monitor} Architecture model and resource metrics. & 
% \cellcolor{Monitor} Estimation and optimization of consumed resources. & 
% \cellcolor{Monitor} Energy consumption reports. \\
% \hline
% %%%%%%%%%%%%%%%%%%%%%%%%%%%%%%%%%%%%%%%%
% % Row 
% Plan Design & Alternative architecture models. & Simulation and comparison of alternative architecture models focusing on horizontal scaling, low baseline resource consumption, and appropriate load balancing techniques. & Simulation results for various architecture variants. \\
% \hline
% %%%%%%%%%%%%%%%%%%%%%%%%%%%%%%%%%%%%%%%%
% % Row 
% Plan & Architecture model. & Scenario-based reliability analysis of the architecture. & Report on the software's reliability. \\
% \hline
%%%%%%%%%%%%%%%%%%%%%%%%%%%%%%%%%%%%%%%%
% Row 
\pmonitor & 
\cellcolor{Monitor} User click behavior within an already deployed UI. & 
\cellcolor{Monitor} Comparison with learned database of personas \cite{zhang2016data}. &
\cellcolor{Monitor} Suggestions for UI improvements (potential future work). \\
\hline

\end{tabular}

\label{tbl:EVA}
\end{table*}
\egroup

With such data from the engineering tools, models about the process, stakeholders, and intended software product, and services in place, a comprehensive DT can support software engineering processes across the complete software lifecycle. 
The next section illustrates such a DT focusing on the creation of sustainable software.

\section{Example Application}
\label{sec:Example}

When creating a DT for software engineering, we must first define the purpose for which we aim to develop it. 
Each of the purposes determines, which \textit{data} is needed from which tooling in the software engineering process, which \textit{process-related models} should be reused or have to be developed, which \textit{services} should be developed to fulfill the purpose, e.g., analysis, optimization, prediction of specific aspects of the software and its related processes, and which \textit{visualizations and interaction possibilities} are needed for human users of this DT. 
Regarding the sustainability of software systems, a similar approach has analyzed which data, models, and services could be of interest when aiming to perform a sustainability analysis of a software system as a purpose~\cite{HHMR23}. To realize such a system, a domain and purpose-tailored specification of data, models, and services is needed, as we show in the following two examples.

% example 1 - more high level
Imagine creating a DT of a software engineering process with the purpose of \textit{analyzing the adherence to the architectural specification of the system under development}. 
Required data and models could be, e.g., a wiki documenting architectural descriptions, and architecture design models as well as the code of the current implementation. 
Automated services analyze the code for potential contradictions with the architecture description, e.g., logically separate components unexpectedly communicate, and provide a high-level overview about which part of the specified architecture was already realized.
The visualization of the analysis results depends on whom to show them: 
For software architects, it might be more interesting to show a visual representation of the software architecture, e.g., as an architecture model or a graph structure, with highlights where contradictions occur and the possibility to either jump into relevant parts of the implementation or which developers to contact to discuss this deviation with. 
For software developers, it might be more interesting to get information directly in the IDE to not distract them from the implementation flow by navigating to another tool.
Here, relevant parts of the architectural specification could be shown, e.g., in components that communicate unexpectedly with another component, with the possibility of retrieving more details. 

In another example, we aim to develop a DT of a software engineering process with the purpose of better ticket planning support for developers. 
Required data are the assigned issues with the estimated effort until the next sprint, the incoming and confirmed new appointments in the developer's calendar and the tickets that the developer closes. 	
Automated services calculate whether there is enough time left until the end of the sprint to process the remaining open tickets. 
If the free time is insufficient, the developer and/or team leader could be warned via email or again in their DT cockpit where such information is collected. 

% we can build DTs fast
Creating dashboards for DTs can be highly automated, e.g., with generators for web applications~\cite{BGK+24}. One can generate large parts of these user interfaces based on data models describing the data to be displayed and models describing the graphical user interfaces. In addition, there exist libraries with configurable GUI components~\cite{GJMR24} to create such dashboards. 
An example DT dashboard for visualizing parameters in a system engineering process for wind turbines can be found in~\cite{MNN+22}. The authors describe the extraction of data from external engineering tools as well as data visualization. Similar approaches can be applied to software engineering processes.

% Most challenging aspects: connect to existing tooling

\section{Discussion}
\label{sec:Discussion}

%\emph{Observability.} 
Not all actions, decisions, and changes to a SE process might be digitally observable easily or at all.  
This includes anything discussed in person without providing protocols for this discussion, as well as business decisions affecting the project(s) in question, such as reducing the developer time assigned to them. 
Also, some data sources that might be useful to optimize the overall process might not be usable due to regulatory or compliance issues (e.g., scanning chat logs for issue patterns might be in conflict with the GDPR).
For this, aggregating and abstraction required information locally on the developers' computers and making transparent what is shared with the DT might improve support for that system. 

%\emph{Tool connection}. 
As the data will be spread over different applications, mechanisms to extract relevant information from each of them in an automated way have to be developed. Here, vendor lock-in and lacking APIs to reach the information in an automated way might be an additional challenge. 

%\emph{Annoyingness}. 
Often, assistance systems that interact with users directly, must be fine-tuned to make and suggest changes as often as necessary, without bothering the users as this can lead to information fatigue, and ultimately, the users may ignore provided information or deactivate the assistance system.
Likewise, information overhead must be prevented. 
Both of these require empirical studies on the application of suggestions and artifact changes through the DT during SE projects.

\section{Future Plans}
\label{sec:FuturePlans}

\emph{Prototype.} We are currently developing DTs for cyber-physical systems in various contexts and started developing sensors and actuators for interacting with software engineering processes. 
These include interfacing with GitHub in its role as a project management tool and CI/CD pipeline as well as monitoring of project schedules and developer capacities.
To start experimenting with these, we plan to create an extension for Visual Studio Code that captures these to (a) suggest improvements to the and (b) make changes to code artifacts directly. 
We will use this as the basis for empirically investigating the benefits of different services in the SE process DTs. 

\emph{Pilot studies.} Our institutes comprise app. $100$ people who are developing software on a daily basis in different domains and in a variety of project contexts and sizes.
Consequently, we will deploy the first prototype into the working groups of our institutes to evaluate and improve them in these contexts.   

\emph{Empirical evaluation.} Once the improved prototype is ready for experimentation, we will investigate its usability regarding correctness, completeness, and helpfulness of suggestions together with user experience experts. 
These experiments will be conducted with novice developers (lab courses with students of different semesters) and experienced developers (from our research transfer networks and in large consortium projects). 

\emph{Functional extension.} In parallel, we plan to prioritize the development of additional services of the DT with focus groups on different phases of the DevOps lifecycle and from different domains, including at least manufacturing, robotics, and construction.

\emph{Systems engineering.} Supporting engineering processes with DTs does not need to be constrained to software processes.  
Some, especially organizational decisions, can be supported by DTs in the same way. 
However, for every reasoning about the subject matter, corresponding services must exist, which, in systems engineering, might pose challenges regarding the availability of data, models, and best practices.
In the future, we, therefore, plan to experiment with DT support in systems engineering projects as well. 
Therefore, we will leverage systems engineering projects in which we are involved with companies in automotive engineering, construction, and machine engineering.

\section{Conclusion}
\label{sec:Conclusion}

% Novelty: The novelty and innovativeness of contributed solutions, problem formulations, methodologies, and/or theories, i.e., the extent to which the paper is sufficiently original with respect to the state of the art
We presented a vision for DTs of software engineering processes that capture not only the typical data and models used today but can leverage additional data (e.g., schedules) and models (such as traces or architecture models) for a novel, holistic,  representation of and action on a specific software engineering process.
% Relevance: The relevance of the research to the field of software engineering
Such holistic software engineering DTs will significantly improve the way we create, maintain, and operate software by providing novel insights and immediate improvements by combining models of the process with data from yet untapped sources and powerful AI services.
% Impact: The significance and potential impact of the research. The potential of the research to disrupt the current practice
This will not only impact established practices of formally trained software professionals but especially of domain experts who create software or contribute to it by applying automated best practices and established software engineering principles to their solutions. 
Hence, liberating significant software engineering potential that can be applied to innovating and creating added-value despite the lack of skilled software engineers.
Ultimately, such DTs will become a competitive factor in software engineering.

\subsubsection*{Acknowledgments}

The authors were partly funded by the Deutsche Forschungsgemeinschaft (DFG, German Research Foundation) - Model-Based DevOps - 505496753. Website: \url{https://mbdo.github.io}.

The authors of the University of Stuttgart were partly funded by the Ministry of Science, Research and Arts of the Federal State of Baden-Württemberg within the In\-no\-va\-tions\-Cam\-pus Future Mobility (ICM).

The authors would like to thank the German Federal Ministry for Economic Affairs and Climate Action (BMWK) for partly funding this work via the joint project: Factory-X. Website: \url{https://factory-x.org/de/}

%\pagebreak

\balance
\bibliographystyle{IEEEtran}
\bibliography{src/bib/main}

\end{document}